C. Örnek*[1,2,3], R. Fechte-Heinen[1,2]

# Designing Hydrogen Permeation Barriers in Titanium Aluminium Nitride through First Principles Density Functional Theory Calculations

Summary

This study investigates hydrogen permeation in titanium aluminium nitride (TiAlN) using ab initio density functional theory (DFT) for cubic and hexagonal crystal structures. Despite the significance of hydrogen barriers, the potential of TiAlN has not been fully explored. We analyzed site specificity, temperature-dependent insertion, and atomic hydrogen migration path energies. Our research highlights the decisive role of crystallographic structure over chemical composition in designing materials resistant to hydrogen absorption. However, once absorbed, hydrogen diffusion is governed by the local chemical environment. Specifically, hydrogen migration through an Al-N plane requires more energy than through Ti-N, which affects the overall diffusion process. We found hydrogen absorption is highly endothermic, with insertion energies from 50 to 320 kJ/mol of hydrogen atom, indicating low uptake probability at ambient conditions. Higher temperatures further increase the energy required, making absorption less favourable. We also identified substantial energy barriers for hydrogen migration, particularly in the hexagonal structure, with peaks up to 276 kJ/mol, indicating a very low probability of migration. These findings underscore TiAlN's exceptional resistance to hydrogen permeation, making it a promising candidate for high-pressure applications such as hydrogen storage and nuclear reactors.

**\*Corresponding author: *Dr. Cem Örnek*,** E-Mail: oernek@iwt-bremen.de, Tel.: +49 421 218 51369

**Affiliations:**

[1]Leibniz-Institut für Werkstofforientierte Technologien – IWT, Badgasteiner Str. 3, 28359 Bremen

[2]MAPEX Center for Materials and Processes, Universität Bremen, Postfach 330 440, 28334 Bremen

[3]Istanbul Technical University, Department of Metallurgical and Materials Engineering, 34469 Istanbul

# 1 Introduction

The urgency to address global warming and the energy crisis drives the shift towards a hydrogen-based energy system [1, 2]. For instance, Germany and the EU aim for carbon neutrality by mid-century [2], transitioning from nuclear and fossil fuels to renewable sources like wind and solar power [3]. Hydrogen energy from renewable sources offers a sustainable, carbon-neutral alternative. However, it presents technical challenges to materials, such as safety and costs. The supply chain for material availability with the technical specifications at low expenses needs to be guaranteed for an efficient energy transition. The vital technological challenge to materials is the possible mechanical fracture due to hydrogen-induced degradation, which often compromises their structural integrity or the risks for leakage [4-8]. Hence, practical hydrogen permeation barriers are essential to prevent leaks in fuel tanks and pipelines, ensuring the integrity and safety of hydrogen infrastructure. Permeation or absorption barriers are vital for materials susceptible to hydrogen embrittlement.

In this regard, permeation barriers with well-known mechanical and electrochemical properties become essential. Herein, the practical significance of TiAlN coatings extends to critical industries such as hydrogen storage and nuclear reactors, where the prevention of hydrogen permeation is crucial in maintaining system integrity. This study aims to bridge the gap between computational predictions and real-world applications, offering insights that could lead to the development of more effective hydrogen barriers. By understanding the atomic-scale interactions within TiAlN, we can optimize coatings to meet the rigorous demands in applications, ultimately contributing to safer and more efficient hydrogen technologies.

Austenitic stainless steels, particularly Types 316 and 316L, are widely used in hydrogen service applications for their corrosion resistance and high-temperature strength. Despite their resistance to hydrogen-assisted fracture, 316 stainless steel can still permeate hydrogen, underscoring the need for permeation barriers. These barriers prevent hydrogen ingress that can compromise material integrity, especially at low temperatures required for hydrogen energy systems [9]. Ceramic coatings, such as alumina and titanium-based ceramics, show promise in reducing hydrogen permeability but face challenges like peeling due to thermal expansion differences and unclear relationships between permeation mechanisms and coating microstructure [10]. Alumina has a high permeation-reduction factor, though its practical application is still evolving [10]. Titanium ceramic coatings (e.g., TiN, TiC, $TiO_2$) have also been explored, with permeation-reduction factors ranging from 10 to over 10,000 [11]. However, the relationship between hydrogen permeation mechanisms and the microstructure of these coatings remains unclear, likely influenced by defect types and numbers.

Transition metal nitrides, particularly titanium aluminium nitride [12], are leading in next-generation hydrogen permeation barrier layers (HPBLs) due to their chemical resistance, ease of fabrication, hardness, and inherent hydrogen resistance [13]. The performance of these materials is closely linked to their microstructural characteristics and chemical composition, necessitating detailed experimental and theoretical exploration. Physical Vapor Deposition (PVD) technologies enable precise control of deposition conditions, tailoring the microstructure to enhance functional properties. TiAlN coatings can be produced with various Ti/Al ratios, leading to a metastable structure [14]. These coatings undergo spinodal decomposition into cubic TiN and hexagonal (h) wurtzite AlN phases after heat exposure to elevated temperatures. The kinetics of this decomposition, especially whether the coating is Al-rich or Al-poor, is not entirely clear and may influence hydrogen permeability [14]. Understanding the different structures and chemistries is essential to grasp the hydrogen permeation characteristics in TiAlN. Notably, a purely hexagonal phase cannot form independently within TiAlN coatings; instead, a combination of cubic TiN and hexagonal AlN must exist, with Al-rich coatings containing more AlN. Detailed information about the structure and properties of TiAlN coatings and further examination of these aspects would be beneficial.

Density functional theory (DFT) simulations are decisive for understanding atomic-scale interactions governing hydrogen diffusion and retention in barrier materials [15]. In this work, we used DFT to study hydrogen interaction with TiAlN in cubic and hexagonal structures, focusing on calculating hydrogen insertion energies as a function of temperature. Our investigation explores potential hydrogen occupation sites within TiAlN, aiming to identify configurations that impede hydrogen mobility to develop better hydrogen barrier coatings. Our study details the potential energy barriers, illuminating hydrogen diffusion kinetics. This study advances the understanding of hydrogen-material interactions for safer hydrogen technologies. Through DFT simulations, we aim to develop barrier technologies to meet safety and efficiency requirements. Insights from this work are expected to impact various fields, offering new possibilities for harnessing hydrogen energy. We identify potential sites within the lattice where hydrogen atoms are absorbed and trapped, excluding grain boundaries, which will be addressed in future work. Future studies will also examine the influence of passivating oxide layers on TiAlN, as these layers may significantly impact hydrogen permeation. We also exclude the spinodal decomposition at high temperatures.

## 2 Computational Methodology

### 2.1 Simulation Environment and Model

We conducted a comprehensive computational study using the MedeA Core 3.8.1 suite, which utilizes VASP 6 (**V**ienna **a**b initio **s**imulation **p**ackage) [16-18] to understand hydrogen absorption and permeation in TiAlN. Our study focused on two crystalline forms of TiAlN, a polymorphic material, cubic (Fm-3m) and hexagonal (P63/mmc). We aimed to identify how hydrogen interacts with the structure, explicitly looking at absorption sites, diffusion pathways, and energy barriers. We used the MedeA 'find empty space' tool to identify vacancies (interstitial lattice sites) for hydrogen insertion, performed structure minimizations to determine the stability of hydrogen-occupied sites, and utilized the Transition State Search (TSS) module to explore hydrogen migration pathways. Hydrogen diffusion barriers were identified through layers of Al-N and Ti-N, and kinetic Monte Carlo simulations were performed to model diffusion paths.

## 2.1 Computational Parameters

For accurate electronic structure calculations, we employed the Generalized Gradient Approximation (GGA) within the Perdew-Burke-Ernzerhof (PBE) functional and set a plane-wave cut-off energy of 500 eV with a k-point density of 0.25 Å$^{-1}$ and electronic convergence criterion of 10$^{-6}$ eV. Geometry optimizations included atom positions, unit cell shape, and volume. Symmetry was reduced to P1 for detailed hydrogen insertion studies. The reader is referred to the Appendix for additional computational parameters, including details on model development.

The insertion energy, also known as formation energy [19], provides insights into the thermodynamic favourability of hydrogen absorption into the structure, with positive values indicating an endothermic process (energy required for insertion). Therefore, the insertion energy ($E_{Insertion}$) was calculated to understand the energetics of hydrogen uptake into the matrix, which involved computing the total energy of the system with a hydrogen atom inserted ($E_{Structure+H}$), the pristine or hydrogen-free structure ($E_{Structure}$), and an isolated hydrogen atom ($E_H$). The insertion energy was defined as:

$$E_{Insertion} = E_{Structure+H} - (E_{Structure} + E_H) \quad \text{(Equation 1)}$$

Then, we calculated the insertion energy at different temperatures ($E_{Insertion}(T)$) to evaluate how hydrogen interaction varies with temperature. The computation involved calculating the Helmholtz free energy for the system with hydrogen ($F_{Structure+H}(T)$), the pristine structure ($F_{Structure}(T)$), and an isolated hydrogen atom ($F_H(T)$). The temperature-dependent insertion energy is given by:

$$E_{Insertion}(T) = F_{Structure+H}(T) - [F_{Structure}(T) + F_H(T)] \quad \text{(Equation 2)}$$

Phonon calculations were conducted to include vibrational contributions to the free energy at various temperatures, ensuring a comprehensive thermodynamic assessment. The calculations have been performed with MedeA Phonon using PHONON Software 6.14 © [20].

Finally, we used the Nudged Elastic Band (NEB) method to study hydrogen diffusion. The NEB method identifies the minimum energy path (MEP) and transition states between adjacent vacancies. The migration energy ($E_{migration}$) was calculated as:

$$E_{migration} = E_{transition\,state} - E_{initial\,state} \quad \text{(Equation 3)}$$

The migration energy is the energy required for a hydrogen atom to move between stable sites in the lattice. Our methodology provides critical insights into hydrogen stability, energetics, and migration barriers in the structure, which is essential for predicting the material's performance in hydrogen-rich environments. The reader is referred to the appendix for a detailed explanation of our computational methodology, including the simulation environment, model structures, and specific parameters used in our DFT calculations and transition state searches.

# 3 Results

## 3.1 Site specificity for hydrogen insertion

The calculated electronic energies in the ground state for the cubic structure, with a hydrogen atom inserted into possible interstitial sites, showed no variation. The cubic structure has eight voids with an equivalent electronic energy of -68.07 eV (-6568.03 kJ/mol) due to the structural symmetry of the cubic lattice. This energy is the total energy of the lattice, Ti$_2$Al$_2$N$_4$H, including all atoms with hydrogen. The total energy for the lattice ((TiAlN$_2$)$_2$) without hydrogen was calculated to be -67.00 eV (-6464.79 kJ/mol). The lower energy with hydrogen is due to the hydrogen atom forming additional bonds within the lattice, stabilizing the structure. The formation of these bonds releases energy, resulting in a lower total energy for the hydrogen-containing lattice than the hydrogen-free lattice. The provided information does not determine whether the insertion energy is endothermic or exothermic. Using equation 1 and considering the energy for one hydrogen atom calculated to be -3.38 eV (-326.23 kJ/mol), an insertion energy of 2.31 eV (222.88 kJ/mol) was determined, indicating that adding one hydrogen is an endothermic and highly energetic process.

**Figure 1** shows the calculated electronic and insertion energies in the ground state for the hexagonal structure with hydrogen atoms in various interstitial voids. The electronic energy without hydrogen is -143.81 eV (-13875.32 kJ/mol) per cell $(Ti_4AlN_3)_2$. This energy differs significantly from the cubic lattice structure due to the different number of atoms per cell. Specific void sites influence energy values, with voids 13-17 and 22-24 showing higher energy levels, suggesting significant impacts on electronic structure and stability when hydrogen is incorporated. This variability indicates differences in hydrogen affinity and electronic properties, affecting material performance. Insertion energies reveal potential hydrogen absorbability in hexagonal TiAlN compared to the cubic structure. While hydrogen absorption in hexagonal TiAlN is endothermic, there is a strong likelihood of absorption into low-energy lattice void sites. These specific sites will be further analyzed to calculate hydrogen diffusion energy barriers and better understand the diffusion process. The high insertion energies calculated for both structures indicate a solid resistance to hydrogen uptake, which is crucial for preventing hydrogen permeation and, thus, embrittlement in materials used for hydrogen storage and transportation. In practical terms, TiAlN coatings could significantly enhance the durability and safety of hydrogen storage systems by minimizing the risk of hydrogen-induced material degradation.

## 3.2 Temperature-dependent insertion energies of hydrogen

The graph presented in **Figure 2** illustrates the calculated insertion energy of hydrogen across the host matrices of TiAlN as a function of temperature, ranging from zero to 3000 K. The calculation results show that the insertion energies of hydrogen are positive, indicating that hydrogen incorporation is generally endothermic, representing a barrier to hydrogen diffusion, increasing with temperature. The hexagonal structure with H in voids 12 or 13 exhibits moderate hydrogen insertion energy profiles for two distinct hydrogen accommodation sites, far lower than for the cubic structure. The similarity indicates that the choice of accommodation site has minimal impact on the overall hydrogen insertion energy. Instead, the crystal structure and local chemical composition dominate in determining hydrogen absorption behaviour and potentially influence diffusion kinetics.

The increasing trend in insertion energy with increasing temperature indicates that higher temperatures lead to higher energy requirements for hydrogen uptake, possibly due to enhanced lattice vibrations and increased activation barriers for hydrogen diffusion. This observation can be attributed to the increased kinetic energy of the atoms in the lattice at higher temperatures, leading to more vigorous lattice vibrations. These vibrations create a dynamic and potentially more disruptive environment for hydrogen atoms, resulting in higher energy barriers for diffusion despite the overall lattice expansion. The results indicate that the activation barrier for hydrogen diffusion, the energy required to move hydrogen atoms from one site to another within the lattice, also increases with temperature due to the need to navigate through a more vibrationally active lattice. The complex energy landscape at higher temperatures, with more local minima and maxima, further contributes to the increased energy requirements for hydrogen uptake.

It should be noted that the Gorsky effect and physical intuition may suggest that increased temperature should facilitate hydrogen diffusion [21]. Higher temperatures cause lattice expansion and increase the kinetic energy, theoretically lowering the energy barriers for hydrogen uptake because more significant interstitial sites should easily accommodate hydrogen atoms, reducing static insertion barriers. The Arrhenius relation, which describes the temperature dependence of diffusion coefficients, supports this idea by indicating that diffusion rates typically increase exponentially with temperature. Studies have shown that hydrogen diffusion coefficients in metals exhibit such temperature dependence as measured through the Gorsky effect. For example, research on the diffusion of hydrogen in niobium and tantalum has found that diffusion coefficients follow Arrhenius-type behaviour, with higher temperatures leading to faster diffusion rates [21].

However, this effect is not seen in our calculation results, most likely due to static and dynamic impact governing the interactions of hydrogen with the lattice atoms, including the influence of quantum fluctuations and anharmonic lattice vibrations, which are not fully captured by the Gorsky effect [22]. While lattice expansion can reduce static barriers, enhanced lattice vibrations at higher temperatures create temporary local distortions that hydrogen atoms must navigate, increasing practical energy barriers despite overall lattice expansion. Additionally, the total free energy of the system includes contributions from both enthalpy (related to lattice expansion) and entropy (related to lattice vibrations and disorder). At higher temperatures, the entropy contribution becomes significant, potentially offsetting the benefits of lattice expansion and increasing the practical energy barriers, leading to increased insertion energy with temperature.

### 3.3 Migration energies

**Figure 3** presents the MEP profiles for hydrogen migration in the cubic structure along three different crystallographic planes: (010), (100), and (001). The hydrogen migration through the (010) plane in **Figure 3(a)** shows the highest energies, indicating the rate-determining path for hydrogen diffusion along voids 7 – 8 – 23. The energy profile shows that the highest energy barrier, 118 kJ/mol, occurs between voids 7 and 8, followed by a lower barrier of 73 kJ/mol between voids 8 and 23, indicating a less favourable migration pathway in this direction. **Figure 3(b)** presents the migration path energies along voids 8 – 36 – 40 through the (100) plane. Here, the energy barriers are relatively moderate, with a peak energy barrier of 23 kJ/mol occurring at voids 8 and 36 and a similar barrier between voids 36 and 40, indicating a more favourable migration pathway than the (010) plane. It should be noted that the voids 8, 36, and 40 were not sites with the least energy minimum, as seen by the four energy minima in the MEP profile in **Figure 3(b)**. The energy minima were about 11 kJ/mol, resulting in a net migration energy barrier of 34 kJ/mol. **Figure 3(c)** illustrates the migration path along voids 8 – 14 – 16 across the (001) plane. The energy barriers in this plane are also moderate, with the highest barrier of 23 kJ/mol between voids 8 and 14 and a slightly lower barrier of 20 kJ/mol between voids 14 and 16. This plane, similar to the (100) plane, suggests favourable migration paths with relatively low energy barriers.

**Figure 4** shows the MEP profiles for hydrogen migration in the hexagonal structure, highlighting a more complex migration route than in the cubic structure. We identified several migration routes but showed the ones along the c-axis only, i.e., the [001] direction. The migration path involves several voids: 13, 9, 10, or directly from 19 to 10, then 8, 7, 12, 19, and 17, with the arrows indicating the direction in which hydrogen atoms migrate through these voids, as seen in **Figure 4(a)**. While the calculations were performed one way, the MEP profiles are also relevant in the backward direction. The MEP profile in **Figure 4(b)** shows the energy barriers hydrogen atoms encounter along this path. The migration energies are symmetric, indicating reasonable confidence in the calculation model. The highest energy barriers in this path are between voids 13 and 9 and voids 19 and 17, which suggests a challenging step for vacancy migration.

**Figure 4(c)** provides a bar graph illustrating the energy barriers associated with each migration path, quantifying the energy peaks and valleys encountered. The energy barriers were obtained by performing structural minimization calculations of the found transition states. The most significant energy barriers are between paths 12–17 and 19–17, with energy values reaching up to 244 kJ/mol and 276 kJ/mol, respectively. Other paths, such as 8–9, 10–8, 7–8, and 19–12, exhibit much lower energy barriers, indicating that these pathway parts are more accessible for hydrogen to navigate. The highest energy barrier for hydrogen diffusion was seen when passing through a stacked layer of nitrogen and titanium, showing the highest barrier effect than all other atomic configurations within the lattice. Overall, the widely varying energy barriers demonstrate the heterogeneity in the hydrogen migration landscape. The substantial energy barriers for hydrogen migration observed in the hexagonal structure underscore the potential of TiAlN as a robust hydrogen permeation barrier, particularly relevant for high-pressure hydrogen environments, such as in hydrogen fuel storage tanks and reactors, where the prevention of hydrogen diffusion is essential for maintaining system safety and longevity.

# 4 Discussion

The calculations have shown that hydrogen absorption into titanium aluminium nitride is a highly endothermic process that necessitates extensive energy independent of the structure, position, and local chemistry, demonstrating its barrier properties. Insertion energies ranging from 50 to 320 kJ/mol suggest a very low likelihood of hydrogen uptake in these structures, as hydrogen infusion is thermodynamically unfavourable under ambient conditions. High endothermic energies require significant thermal or chemical energy input to overcome barriers. Although higher temperatures could increase diffusion rates, hydrogen uptake remains unlikely without additional facilitating mechanisms. Materials with higher positive insertion energies can act as practical barriers to hydrogen permeation, making them less prone to saturation and ideal for applications like protective cladding in nuclear reactors or containers for hydrogen transport and storage. However, high insertion energies might also indicate susceptibility to hydrogen embrittlement, where hydrogen accumulates at defects, grain boundaries, or interfaces, potentially leading to material failure [23]. High insertion energies mean that hydrogen atoms require significant energy to be absorbed into the bulk lattice.

Hydrogen atoms migrate and accumulate at energetically favourable sites such as defects, grain boundaries, or interfaces, where the local environment provides a lower energy barrier for insertion. This accumulation can cause local stress concentrations, reduce cohesive strength, and enhance diffusion pathways, leading to hydrogen-induced decohesion and phase transformations that degrade mechanical properties [8, 24, 25]. Given the strong affinity of hydrogen to titanium and aluminium, the formation of potential hydrides cannot be disregarded [26]. Additionally, the formation of molecular hydrogen can create internal pressure, initiating and propagating microcracks, thus increasing the material's susceptibility to hydrogen embrittlement and failure [27, 28]. Hence, materials with positive insertion energies can be excellent candidates for hydrogen permeation barriers if they possess the necessary mechanical properties and resistance to embrittlement.

The insertion energy for the cubic structure increases at a higher rate, indicating a different interaction of hydrogen with the lattice that is more sensitive to temperature variations, which could reflect a more efficient barrier mechanism for hydrogen uptake. The varying rates at which insertion energy increases for these two structures underscore the influence of crystal structure on the thermodynamics of hydrogen absorption. The subtle nuances in how temperature influences the crystal lattice of these compounds, such as through thermal expansion and vibrational entropy, become critical. Moreover, the effect of temperature on the electronic properties – altering the band gap and the distribution of electronic states – could also modulate the energy required for hydrogen insertion. It should be noted that DFT calculations are grounded in approximations and assumptions that model the behaviour of electrons within a material's lattice. These calculations, while insightful, are an idealization that needs to be validated against experimental data aimed at future work. The consistent increase in energy with temperature for both materials, despite their different structures, points toward a common thermodynamic principle governing hydrogen insertion.

To clarify endothermic energy, we give an example showing its practical impact. An insertion energy of 50 kJ/mol for hydrogen signifies a very high energy barrier that hydrogen atoms must overcome to enter the material, reflecting a high resistance to hydrogen permeation. The thermal energy of hydrogen atoms in a solid matrix such as titanium aluminium nitride can be estimated using principles from statistical mechanics. When trapped in the lattice, each hydrogen atom can have three degrees of freedom corresponding to its vibrations in three-dimensional space. According to the equipartition theorem, each degree of freedom contributes $\frac{1}{2} \times k_B T$ to the thermal energy, where $k_B$ is the Boltzmann constant, and T is the absolute temperature. The total thermal energy can be expressed as

$$E_{thermal} = 3 \times \frac{1}{2} k_B T \text{ (Equation 4)}$$

for a hydrogen atom in its gaseous aggregate state. However, when considering solids and their vibrational modes (phonons), a more detailed approach would involve the total vibrational energy, which approximates $3 \times k_B T$ per atom at high temperatures due to kinetic and potential energy contributions from these vibrations. At room temperature (around 298 K), the thermal energy available to hydrogen atoms is approximately 7.4 kJ/mol, significantly lower than the calculated insertion energies, suggesting a very low likelihood of hydrogen permeation in TiAlN under ambient conditions. As temperature increases, so does the thermal energy. For instance, at 500 K, the thermal energy is around 12.5 kJ/mol; at 800 K, it reaches about 20 kJ/mol.

Although these values are still below 50 kJ/mol, the exponential nature of the Arrhenius equation means that even small increases in thermal energy can significantly enhance the (minor) fraction of hydrogen atoms to overcome the barrier. As the temperature rises, the insertion energy effectively increases, making conditions less favourable for hydrogen uptake. Higher temperatures also raise the vibrational entropy of the host lattice and hydrogen atoms, further discouraging hydrogen insertion. However, the energy for hydrogen entry is not solely dictated by the temperature. Pressure or fugacity also has a contributory role, amplified in high-pressure environments, such as hydrogen storage tanks or high-pressure reactors, making hydrogen permeation more likely. The chemical potential drives hydrogen to enter a structure, which is high when the hydrogen pressure (in gas) or fugacity (in electrolyte) is high. For example, at a pressure of 10 MPa (100 bar), the energy change associated with this pressure increase can be estimated using the following relation:

$$\Delta G = RT \ln \frac{P}{P_0} \text{ (Equation 5)}$$

R is the gas constant, T is the temperature, P is the pressure, and $P_0$ is the reference pressure. Assuming room temperature (298 K) and a pressure of 10 MPa, ΔG of ≈5.7 kJ/mol is obtained. So, even at 800 K and a pressure of 10 MPa, the insertion energy of 50 kJ/mol is unlikely to overcome, demonstrating TiAlN's effectiveness as a hydrogen permeation barrier system. However, it should be noted that surface contamination can dramatically alter the behaviour of hydrogen interaction with TiAlN. Contaminants like palladium, platinum, or iron can facilitate hydrogen dissociation and absorption, lowering the effective barrier energy. These elements can create localized sites where hydrogen atoms can more easily dissociate and penetrate the material. Catalytic surface contaminants can effectively reduce the insertion energy in localized regions and may provoke hydrogen uptake. The overall absorption barrier efficiency could be compromised, especially in high contaminant concentrations.

Furthermore, structural inhomogeneities, such as defects like grain boundaries or chemical impurities, may also compromise the ability to overcome the high barrier energy, which may facilitate diffusion to a certain extent. Typically, grain boundaries are active sink sites for hydrogen, potentially supporting hydrogen entry. Nevertheless, in both structural forms, our calculations demonstrate the high expected barrier performance of hydrogen permeation for titanium aluminium nitride. It should also be noted that TiAlN undergoes a spinodal decomposition at elevated temperatures, forming TiN and AlN. At that moment, the diffusion behaviour may compromised.

The detailed migration path analyses presented in **Figure 3** and **Figure 4** provide a comprehensive understanding of hydrogen diffusion in cubic and hexagonal TiAlN structures. The data reveal significant differences in energy barriers across various crystallographic planes and pathways, which are critical for evaluating the practical effectiveness of TiAlN. In the cubic structure, the (010) plane shows notably high energy barriers for hydrogen migration, with the peak value reaching 118 kJ/mol. This substantial barrier indicates that hydrogen diffusion along this path is highly unfavourable, confirming the strong resistance of the (010) plane against hydrogen permeation. Even though the (100) and (001) planes exhibit lower barriers, with peak values of 23 kJ/mol and 20 kJ/mol, respectively, these energy levels are still sufficiently high to deter significant hydrogen migration under typical operating conditions. The lower barriers in these planes suggest that, while hydrogen might find more accessible migration paths than the (010) plane, the overall resistance remains robust, making the cubic TiAlN structure an effective barrier.

The hexagonal structure presents an even more complex and varied migration landscape. The highest energy barriers in this structure reached up to 276 kJ/mol. These extreme values highlight the exceptional resistance of the hexagonal structure to hydrogen permeation. As illustrated by the numerous void transitions (**Figure 3** and **Figure 4**), the complexity of the migration pathways suggests that while some paths may offer lower resistance, significant energetic obstacles hinder the overall diffusion process. This heterogeneity in energy barriers emphasizes the intricate nature of hydrogen migration within the hexagonal structure, supporting its effectiveness as a hydrogen barrier. Overall, the high energy barriers and the complex migration pathways revealed by our calculations confirm that titanium aluminium nitride, in its cubic and hexagonal forms, is a promising material for hydrogen permeation barriers. The significant resistance to hydrogen diffusion, combined with the material's robustness under varying environmental conditions, makes TiAlN an ideal candidate for protective cladding in nuclear reactors, containers for hydrogen transport, and storage systems. These findings underscore the potential of TiAlN to provide reliable and durable solutions in the field of hydrogen containment.

*Implications of Defects and Electronic Structure on Hydrogen Permeation in TiAlN*

While our study provides critical insights into the hydrogen permeation properties of titanium aluminium nitride in its cubic and hexagonal forms, the implications of defect chemistry and electronic structure should also be considered [29]. Our focus has been on defect-free structures, where we identified significant energy barriers for hydrogen absorption and migration, underscoring its potential as an effective hydrogen barrier material [30]. However, unlike metals, nitrides exhibit complex behaviour due to their possible band gap and the role of defects, which can profoundly impact hydrogen transport [29]. In semiconducting or insulating nitrides, defects such as nitrogen vacancies, interstitials, or antisite defects can introduce band gap states, altering the material's electronic properties and influencing hydrogen behaviour [31]. These defects can serve as traps for hydrogen atoms or create alternative migration pathways, which may lower or raise the energy barriers for diffusion [30]. For instance, TiAlN nitrogen vacancies could act as hydrogen trap sites or facilitate easier migration than interstitial sites [29]. This scenario could change the hydrogen migration barriers, contrary to what we observe in defect-free structures.

Moreover, the formation energy of defects is highly sensitive to external conditions such as nitrogen partial pressure, temperature, and the Fermi level [31]. These variables can further influence hydrogen diffusion dynamics, making the permeation behaviour context-dependent. A band gap in TiAlN, if any, implies that the chemical potential and defect formation energies will vary with the Fermi level, particularly under different environmental or processing conditions [29, 31]. The Fermi level, relative to the valence and conduction bands, can significantly affect the formation energies of charged defects, influencing the overall hydrogen absorption and migration behaviour [29, 31]. In metals, such considerations are often negligible due to the lack of a significant band gap. However, this electronic structure effect in nitrides means that the defect formation energies and, consequently, the hydrogen permeation properties can vary depending on the Fermi level position, which may shift under different doping or environmental conditions [31]. Although TiAlN exhibits metallic character (see the appendix for computational evidence) due to strong metal-nitride bonding, introducing defects could locally modify its electronic structure and, consequently, its hydrogen permeation properties. It is also essential to consider that hydrogen uptake generally increases the conductivity of semiconductor oxides [4, 32-34], potentially diminishing any possible band gap. Therefore, while TiAlN's intrinsic properties are crucial, its defect landscape—modifiable through processing conditions—will likely play a pivotal role in determining its overall resistance to hydrogen permeation.

Given these considerations, the design of hydrogen permeation barriers using TiAlN should account for the potential influence of defects and the electronic structure. While our study highlights the intrinsic properties of defect-free TiAlN, future designs should consider strategies for controlling defect concentrations, such as optimizing nitrogen partial pressure during synthesis or introducing dopants that stabilize desirable defect configurations. This approach could enhance the material's resistance to hydrogen permeation under various operational conditions, making TiAlN a more robust material for hydrogen storage and nuclear applications.

# 5 Conclusions

This study investigated the hydrogen permeation properties of titanium aluminium nitride in cubic and hexagonal structures using density functional theory calculations. The results indicate hydrogen absorption into either structure is highly endothermic, with insertion energies ranging from 50 to 320 kJ/mol. These high insertion energies suggest a low likelihood of hydrogen uptake under ambient conditions. The temperature-dependent analysis revealed that higher temperatures increase the energy required for hydrogen incorporation, making the absorption process even less favourable. Migration path analyses showed significant energy barriers for hydrogen diffusion in both cubic and hexagonal structures. In the cubic structure, the (010) plane exhibited the highest barrier at 118 kJ/mol, while the (100) and (001) planes had lower but still substantial barriers. The hexagonal structure presented even higher barriers, with peaks up to 276 kJ/mol, indicating exceptional resistance to hydrogen permeation. These findings underscore the potential of TiAlN as an effective hydrogen permeation barrier suitable for high-pressure environments such as hydrogen storage tanks and reactors. The practical implications of this study highlight titanium aluminium nitride in cubic and hexagonal form as a promising candidate for hydrogen permeation barriers in critical industrial applications, mainly when applied as a coating. The high resistance to hydrogen absorption and migration makes TiAlN particularly suitable for use in hydrogen storage systems and protective cladding in nuclear reactors. Future research should focus on optimizing the microstructural properties of TiAlN coatings to enhance their performance in demanding environments.

# 6 Future Works and Outlook

Understanding the exact mechanisms underlying hydrogen permeation in TiAlN requires a comprehensive approach, combining experimental investigations and theoretical modelling to elucidate the changes occurring at the atomic and molecular levels. To validate the computational findings of this research, we plan to fabricate titanium aluminium nitride and its structural derivatives and perform electrolytic hydrogen permeation experiments.

Furthermore, we plan to explore how varying nitrogen partial pressure during synthesis influences the defect landscape and hydrogen permeation properties. Investigating the effect of doping on the electronic structure and defect formation energies could also provide pathways to enhance the hydrogen barrier properties of TiAlN. Identifying dopants that stabilize desirable defect configurations through high-throughput DFT calculations could significantly improve material performance. These experimental efforts will provide crucial insights into the material's real-world performance as a hydrogen barrier. In addition to the experimental work, we aim to extend our computational investigations by considering the effects of point defects, such as vacancies, dopants, and grain boundaries. These defects will likely play a significant role in hydrogen behaviour, potentially altering the diffusion pathways and energy barriers observed in defect-free TiAlN. By integrating these factors into our simulations, we can develop a more accurate and comprehensive understanding of hydrogen permeation (and resistance).

# Acknowledgement

The authors are grateful to Prof. Lucio Colombi Ciacchi, University of Bremen, for providing access to the Lesum Cluster of the Bremen Center for Computational Materials Science (BCCMS), funded by the German Research Foundation (DFG) under the grant number INST 144/506-1 FUGG (project number 457276215). The authors are also grateful to The North-German Supercomputing Alliance (HLRN) in Berlin, Germany, for providing access to the supercomputer premises. Cem Örnek appreciates the fruitful scientific discussions with and valuable support from Dr René Windiks, Materials Design S.A.R.L. and Dr Mikael Christensen, Materials Design S.A.R.L. The authors thank Dr Norbert Riefler for providing access to the high-power server computer at IWT Bremen. Cem Örnek is grateful for valuable discussions with Dr Andreas Mehner, IWT Bremen and Prof. Mustafa Ürgen, Istanbul Technical University.

# References

1. Oshiro, K. and S. Fujimori, *Role of hydrogen-based energy carriers as an alternative option to reduce residual emissions associated with mid-century decarbonization goals.* Applied Energy, 2022. **313**: p. 118803.
2. Seck, G.S., et al., *Hydrogen and the decarbonization of the energy system in europe in 2050: A detailed model-based analysis.* Renewable and Sustainable Energy Reviews, 2022. **167**: p. 112779.
3. *Unsere Energiewende: sicher, sauber, bezahlbar.* 2024   [cited 2024 04/06/2024]; Available from: https://www.bmwk.de/Redaktion/DE/Dossier/energiewende.html.
4. Örnek, C., et al., *Understanding passive film degradation and its effect on hydrogen embrittlement of super duplex stainless steel – Synchrotron X-ray and electrochemical measurements combined with CalPhaD and ab-initio computational studies.* Applied Surface Science, 2023. **628**: p. 157364.
5. Örnek, C., et al., *The causation of hydrogen embrittlement of duplex stainless steel: Phase instability of the austenite phase and ductile-to-brittle transition of the ferrite phase – Synergy between experiments and modelling.* Corrosion Science, 2023. **217**: p. 111140.
6. Örnek, C., et al., *Hydrogen-Induced Micro-Strain Evolution in Super Duplex Stainless Steel—Correlative High-Energy X-Ray Diffraction, Electron Backscattered Diffraction, and Digital Image Correlation.* Frontiers in Materials, 2022. **8**(598).
7. Koyama, M., et al., *Recent progress in microstructural hydrogen mapping in steels: quantification, kinetic analysis, and multi-scale characterization.* Materials Science and Technology, 2017. **33**(13): p. 1481-1496.
8. Barrera, O., et al., *Understanding and mitigating hydrogen embrittlement of steels: a review of experimental, modelling and design progress from atomistic to continuum.* Journal of Materials Science, 2018. **53**(9): p. 6251-6290.
9. Oyewole, O.L., N.I. Nwulu, and E.J. Okampo, *Optimal design of hydrogen-based storage with a hybrid renewable energy system considering economic and environmental uncertainties.* Energy Conversion and Management, 2024. **300**: p. 117991.
10. Dearnley, P.A. and G. Aldrich-Smith, *Corrosion–wear mechanisms of hard coated austenitic 316L stainless steels.* Wear, 2004. **256**(5): p. 491-499.
11. Tamura, M., M. Noma, and M. Yamashita, *Characteristic change of hydrogen permeation in stainless steel plate by BN coating.* Surface and Coatings Technology, 2014. **260**: p. 148-154.
12. Man, B.Y., et al., *Microstructure, oxidation and H2-permeation resistance of TiAlN films deposited by DC magnetron sputtering technique.* Surface and Coatings Technology, 2004. **180-181**: p. 9-14.
13. Rehman, A., et al., *Chemical Interaction of Hydrogen Radicals (H*) with Transition Metal Nitrides.* The Journal of Physical Chemistry C, 2023. **127**(36): p. 17770-17780.
14. Zhou, J., et al., *Phase equilibria, thermodynamics and microstructure simulation of metastable spinodal decomposition in c–Ti1−xAlxN coatings.* Calphad, 2017. **56**: p. 92-101.

15. Rönnebro, E.C.E., R.L. Oelrich, and R.O. Gates, *Recent Advances and Prospects in Design of Hydrogen Permeation Barrier Materials for Energy Applications—A Review.* Molecules, 2022. **27**(19): p. 6528.
16. Kresse, G. and J. Furthmüller, *Efficient iterative schemes for ab initio total-energy calculations using a plane-wave basis set.* Physical Review B, 1996. **54**(16): p. 11169-11186.
17. Kresse, G. and J. Furthmüller, *Efficiency of ab-initio total energy calculations for metals and semiconductors using a plane-wave basis set.* Computational Materials Science, 1996. **6**(1): p. 15-50.
18. Kresse, G. and D. Joubert, *From ultrasoft pseudopotentials to the projector augmented-wave method.* Physical Review B, 1999. **59**(3): p. 1758-1775.
19. Şeşen, B.M., M. Mansoor, and C. Örnek, *Elucidating the dynamics of hydrogen embrittlement in duplex stainless steel.* Corrosion Science, 2023. **225**: p. 111549.
20. Parlinski, K., Z.Q. Li, and Y. Kawazoe, *First-Principles Determination of the Soft Mode in Cubic ${\mathrm{ZrO}}_{2}$.* Physical Review Letters, 1997. **78**(21): p. 4063-4066.
21. Cantelli, R., F.M. Mazzolai, and M. Nuovo, *Internal Friction due to Long-Range Diffusion of Hydrogen in Niobium (Gorsky Effect).* physica status solidi (b), 1969. **34**(2): p. 597-600.
22. Cheng, B., A.T. Paxton, and M. Ceriotti, *Hydrogen Diffusion and Trapping in α-Iron: The Role of Quantum and Anharmonic Fluctuations.* Phys Rev Lett, 2018. **120**(22): p. 225901.
23. de Souza, G.B., et al., *Structural, chemical and tribo-mechanical surface features of Ti and nitrided Ti submitted to hydrogen low energy implantation.* Materials Chemistry and Physics, 2010. **124**(1): p. 443-452.
24. Vlasov, N.M. and I.I. Fedik, *Hydrogen segregation in the area of threefold junctions of grain boundaries.* International Journal of Hydrogen Energy, 2002. **27**(9): p. 921-926.
25. Örnek, C., B.M. Şeşen, and M.K. Ürgen, *Understanding Hydrogen-Induced Strain Localization in Super Duplex Stainless Steel Using Digital Image Correlation Technique.* Metals and Materials International, 2021.
26. Scully, J.R., G.A. Young, and S.W. Smith, *19 - Hydrogen embrittlement of aluminum and aluminum-based alloys*, in *Gaseous Hydrogen Embrittlement of Materials in Energy Technologies*, R.P. Gangloff and B.P. Somerday, Editors. 2012, Woodhead Publishing. p. 707-768.
27. Lynch, S., *Hydrogen embrittlement phenomena and mechanisms.* Corrosion Reviews, 2012. **30**(3-4): p. 105-123.
28. Robertson, I.M., et al., *Hydrogen Embrittlement Understood.* Metallurgical and Materials Transactions B, 2015. **46**(3): p. 1085-1103.
29. Rahman, M.M., et al., *Chemical bonding states and solar selective characteristics of unbalanced magnetron sputtered TixM1−x−yNy films.* RSC Advances, 2016. **6**(43): p. 36373-36383.
30. Somjit, V. and B. Yildiz, *Doping α-Al2O3 to reduce its hydrogen permeability: Thermodynamic assessment of hydrogen defects and solubility from first principles.* Acta Materialia, 2019. **169**: p. 172-183.
31. Czelej, K., et al., *Atomistic Origins of Various Luminescent Centers and n-Type Conductivity in GaN: Exploring the Point Defects Induced by Cr, Mn, and O through an Ab Initio Thermodynamic Approach.* Chemistry of Materials, 2024. **36**(13): p. 6392-6409.
32. Guo, L., et al., *effect of hydrogen on pitting susceptibility of 2507 duplex stainless steel.* Corrosion Science, 2013. **70**: p. 140-144.
33. Li, M., et al., *The mechanism of hydrogen-induced pitting corrosion in duplex stainless steel studied by SKPFM.* Corrosion Science, 2012. **60**: p. 76-81.
34. Guo, L.Q., et al., *effect of hydrogen on semiconductive properties of passive film on ferrite and austenite phases in a duplex stainless steel.* Scientific Reports, 2017. **7**(1): p. 3317.

Figures

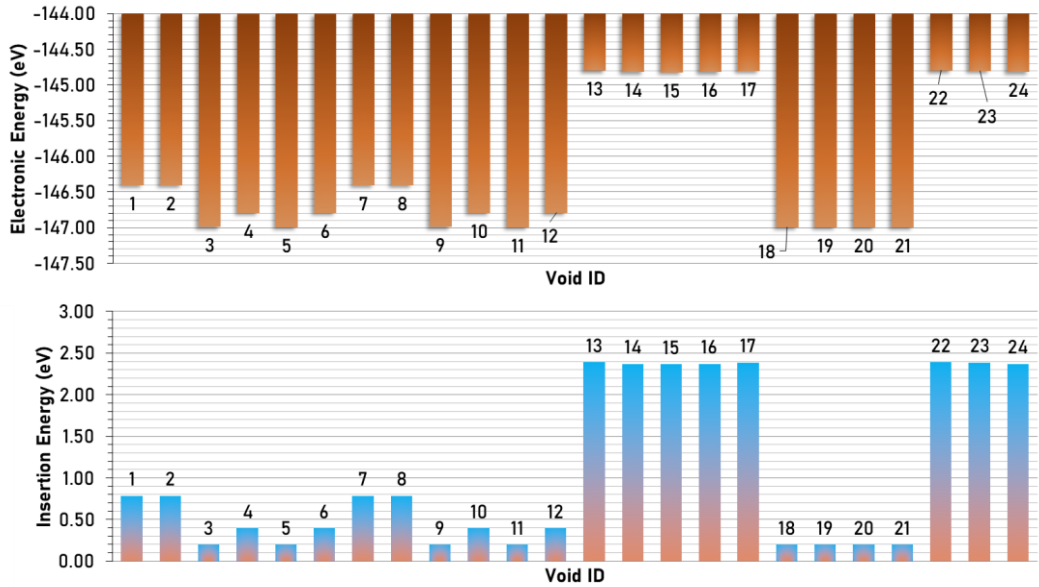

*Figure 1:* Calculated electronic and insertion energies at the ground state of the hexagonal titanium aluminium nitride structure with hydrogen inserted into different interstitial vacancy sites (voids).

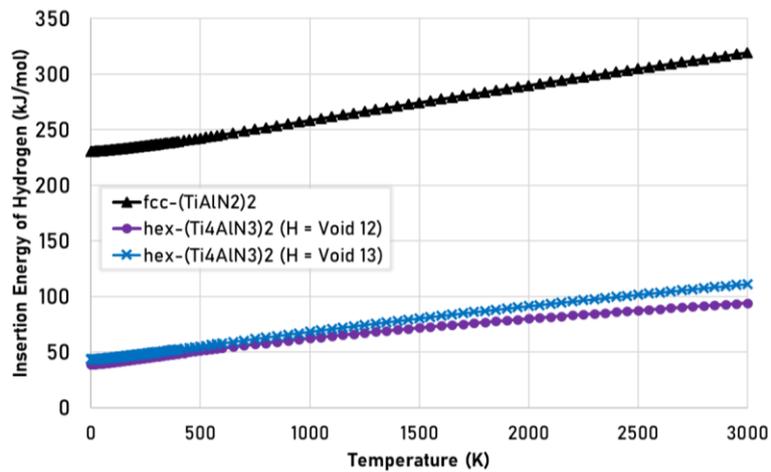

*Figure 2:* The hydrogen insertion energies as a temperature function for titanium aluminium nitride with cubic (fcc) and hexagonal (hex) structure.

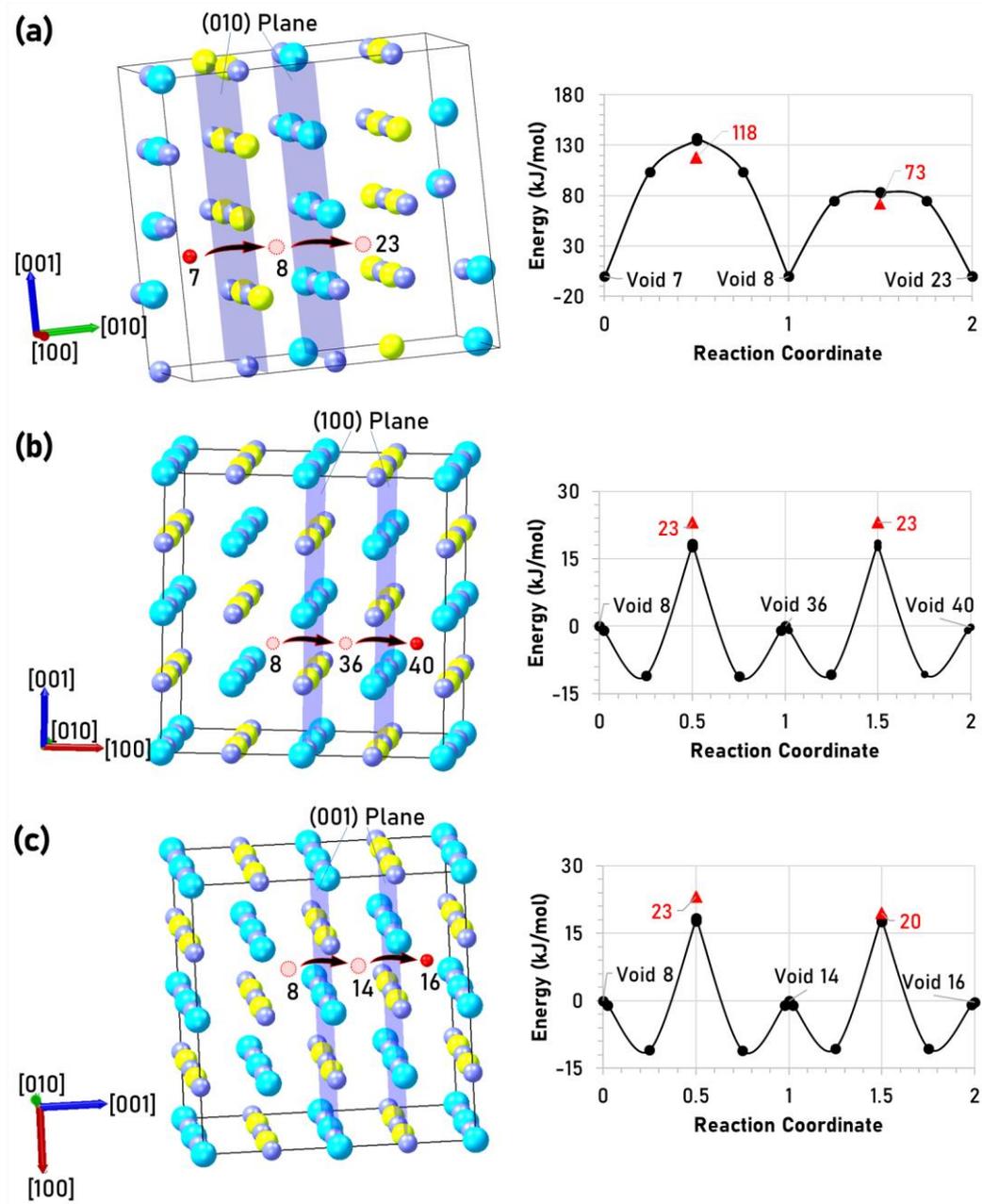

*Figure 3:* DFT-calculated migration path energies in the cubic titanium aluminium nitride structure along the voids (a) 7 – 8 – 23; (b) 8 – 36 – 40; (c) 8 – 14 – 16. The red data points (triangular) show electronic structure calculation with improved accuracy, reflecting the true migration barrier energy. The colour code for the atoms are: titanium is cyan, aluminium is yellow, nitrogen is grey.

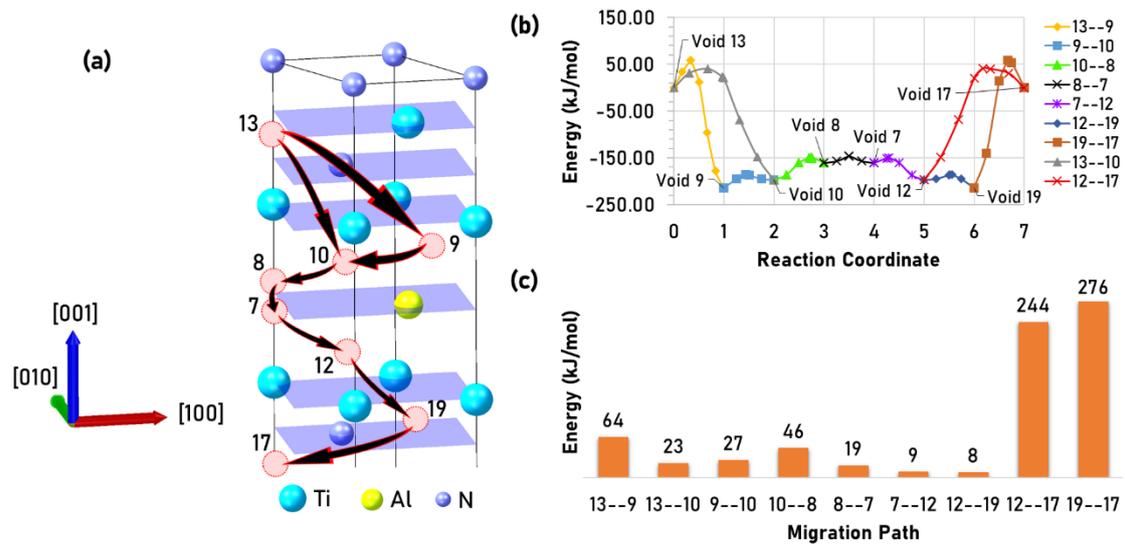

*Figure 4:* (a) Schematic representation of the migration pathways for hydrogen atoms in the hexagonal titanium aluminium nitride crystal lattice, highlighting the various voids and migration paths along the voids 13 – 9 – 10 – 8 – 7 – 12 – 19 – 17. (b) The energy profile for each migration pathway shows the variation in energy (kJ/mol) along the reaction coordinate for different void transitions. (c) Bar graph illustrating the energy barriers associated with each migration path, indicating the most and least energetically favourable pathways. The migration paths are labelled according to the specific transition between voids.

# Appendix

## Computational Methodology

### Simulation Environment

We utilized the MedeA Core 3.8.1 simulation suite for our computational investigations. The Vienna Ab initio Simulation Package (VASP 6) was used for density functional theory calculations of electronic energies. The 'find empty space' feature was used to find vacancies and add a hydrogen atom (defect) to the vacancy. The phonon module was employed to calculate the phonon density of states to obtain the vibrational energy contributions. The Transition State Search (TSS) module was used to identify migration barriers. We aimed to provide essential insights into the hydrogen absorption and permeation characteristics of titanium aluminium nitride relevant to hydrogen storage, diffusion barriers, and related applications. Some calculations were performed at the Lesum Cluster of the Bremen Center for Computational Materials Science (BCCMS) at Bremen University, Germany and The North-German Supercomputing Alliance (HLRN) in Berlin, Germany.

### Model and structures used

Our primary objective was to characterize the energies associated with hydrogen absorption and migration within titanium aluminium nitride, a polymorphic material manifesting in two distinct crystalline forms, with the symmetry space groups Fm-3m and P6_3mc. We sourced the face-centred cubic $TiAlN_2$ (cubic) structural data from the Inorganic Crystal Structure Database (ICSD) under collection code 58012, described by the Hermann-Mauguin symbol Fm-3m, with an empirical formula $(AlTi)N_2$, corresponding to the NaCl structural type. This structure was modified to have an aluminium layer between two titanium layers, as shown in **Figure Appendix 1**.

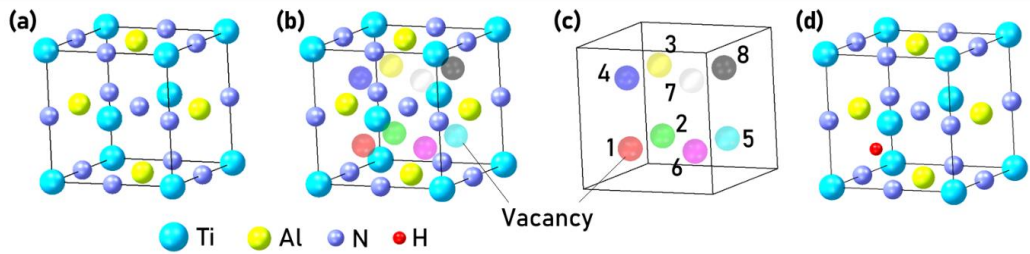

***Figure Appendix 1:*** *(a) The cubic ($TiAlN_2$) structure in the Fm-3m space group showcases Ti atoms (blue), Al atoms (yellow), and N atoms (grey), (b,c) highlighted are interstitial sites; (d) the structure with a hydrogen atom inserted to site 1. Note that the symmetry for the structures in (b-d) was lowered to P1 to enable hydrogen insertion into each interstitial site. Therefore, the vacancies are coloured differently.*

Then, after a structure minimization calculation, the symmetry of the structure was lowered to P1, allowing the insertion of a hydrogen atom into each identified interstitial vacancy. Eight equivalent vacant sites numbered 1 to 8 were identified using the MedeA find empty space tool for possible hydrogen absorption. The positions, atomic radii and fractional coordinates for the vacancy positions are provided in **Table Appendix 1**.

***Table Appendix 1:*** *Fractional coordinates and radii of the vacancies in cubic TiAlN (see also **Figure Appendix 1**).*

| ID | Radius | Coordination | X | Y | Z |
|---|---|---|---|---|---|
| 1 | 0.362 | 2 | 0.2500 | 0.2500 | 0.2500 |
| 2 | 0.362 | 2 | 0.2500 | 0.7500 | 0.2500 |
| 3 | 0.362 | 2 | 0.2500 | 0.2500 | 0.7500 |
| 4 | 0.362 | 2 | 0.2500 | 0.7500 | 0.7500 |
| 5 | 0.362 | 2 | 0.7500 | 0.7500 | 0.2500 |
| 6 | 0.362 | 2 | 0.7500 | 0.2500 | 0.2500 |
| 7 | 0.362 | 2 | 0.7500 | 0.2500 | 0.7500 |
| 8 | 0.362 | 2 | 0.7500 | 0.7500 | 0.7500 |

First, we performed a full structure minimization calculation. Then, we added a hydrogen atom at each vacancy and performed structure minimization to compute the electronic ground state energy. This step was crucial to understand the stability and energy landscape of the system. We then conducted a transition state search using MedeA's TSS module to investigate hydrogen diffusion pathways. A transition state in this context represents the configuration where a hydrogen atom moves from one vacancy to another, overcoming the energy barrier between these sites. The transition state search involves identifying high-energy configurations along the reaction pathway. For example, as shown in **Figure Appendix 1**(d,a) structure where only vacancy 1 is occupied by a hydrogen atom represents a state from which the hydrogen atom can migrate to another vacancy. The transition state search aims to find a saddle point on the potential energy surface. A saddle point is where the energy is at a maximum along the reaction coordinate but at a minimum concerning all other coordinates. Then, this point represents the transition state of hydrogen migration, which is the highest energy state along the reaction pathway between two stable configurations of hydrogen in the lattice. It is important to note that for hydrogen diffusion to occur, some vacancies must remain unoccupied, providing the necessary sites for hydrogen atoms to move into.

Furthermore, we performed kinetic Monte Carlo molecular dynamics to simulate possible hydrogen diffusion paths (not shown). We noted that two energetically distinctive diffusion barriers determine the hydrogen diffusion through the lattice, namely through layers of Al-N and Ti-N. Once the transition states were found, we created a 2x2x2 superlattice (**Figure Appendix 2**) and repeated the analysis for improved computational accuracy. Then, the found transition state was created, and an accurate energy calculation was performed with the same computational parameters as done for energy minimization calculations to be consistent.

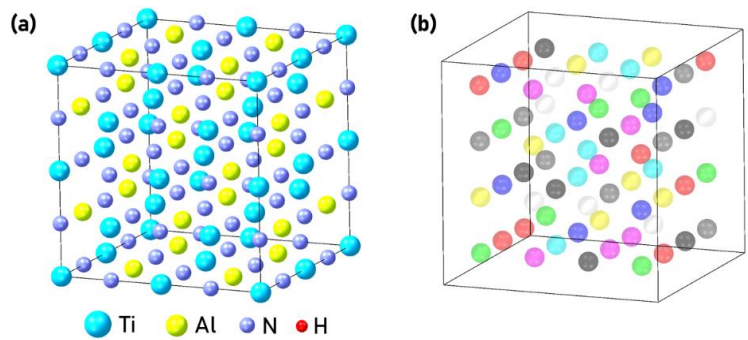

*Figure Appendix 2: (a) Superlattice (2x2x2) of the cubic titanium aluminium nitride structure with 64 atoms created from the unit cell shown in **Figure Appendix 1**; (b) Interstitial vacancy lattice sites for possible hydrogen accommodation. Note that the symmetry for the structure was lowered to P1 to enable hydrogen insertion into each interstitial site. Therefore, the vacancies are coloured differently.*

The structure for the hexagonal variant of titanium aluminium nitride (hex) was obtained from the crystallographic open database (COD) with ID 1526338, space group P63/mmc, with the compositional formula $AlN_3Ti_4$, shown in **Figure Appendix 3**.

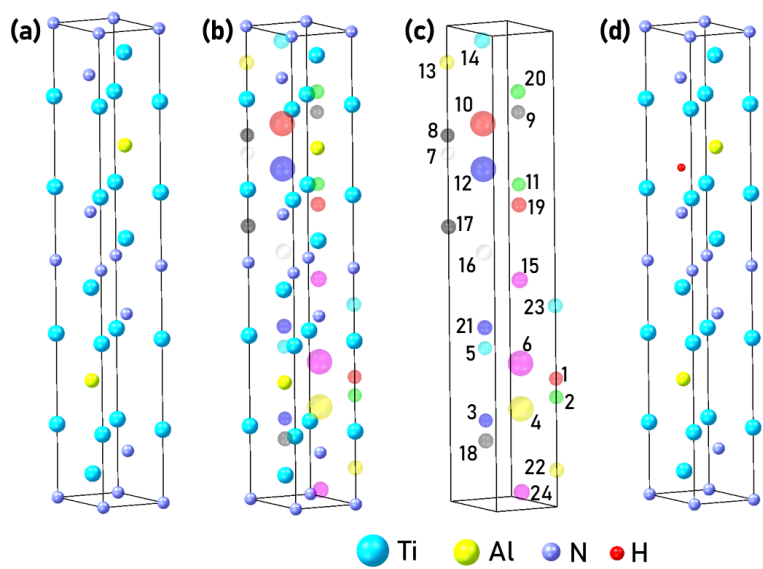

*Figure Appendix 3: (a) The structure of hexagonal titanium aluminium nitride with 24 atoms in the space group P6_3mc; (b) the lattice with empty spaces for possible accommodation sites for atomic hydrogen; (c) the vacancy sites in the motive of the structure numbered from 1 to 24; (d) the lattice with a hydrogen atom inserted to void no. 12. Note that the symmetry for the structures in (b-d) was lowered to P1 to enable hydrogen insertion into each interstitial site. Therefore, the vacancies are coloured differently.*

Likewise, the symmetry of the structure was lowered to P1, allowing identification of each lattice vacancy. The voids are numerically labelled as possible sites for reversible hydrogen trapping. The voids with the smallest radii (see **Table Appendix 2**) represent positions less likely to host hydrogen atoms due to spatial constraints.

*Table Appendix 2: Fractional coordinates and radii of the vacancies in hexagonal TiAlN (see also **Figure Appendix 3**).*

| ID | Radius | Coordination | X | Y | Z |
|---|---|---|---|---|---|
| 1 | 0.338 | 3 | 0.6667 | 0.3333 | 0.2982 |
| 2 | 0.338 | 3 | 0.6667 | 0.3333 | 0.2034 |
| 3 | 0.338 | 1 | 1 | 0 | 0.2315 |
| 4 | 0.611 | 1 | 0.3333 | 0.6667 | 0.1753 |
| 5 | 0.338 | 3 | 1 | 0 | 0.2701 |
| 6 | 0.611 | 3 | 0.3333 | 0.6667 | 0.3263 |
| 7 | 0.338 | 1 | 0.6667 | 0.3333 | 0.8215 |
| 8 | 0.338 | 1 | 0 | 1 | 0.7724 |
| 9 | 0.338 | 3 | 0 | 1 | 0.7257 |
| 10 | 0.611 | 3 | 0.6667 | 0.3333 | 0.6678 |
| 11 | 0.338 | 3 | 0.3333 | 0.6667 | 0.9782 |
| 12 | 0.611 | 3 | 0.6667 | 0.3333 | 0.4714 |
| 13 | 0.361 | 4 | 0.3333 | 0.6667 | 0.5245 |
| 14 | 0.390 | 4 | 0.9976 | 0.0024 | 0.4218 |
| 15 | 0.390 | 4 | 0.3333 | 0.6667 | 0.1333 |
| 16 | 0.390 | 4 | 0.3333 | 0.6667 | 0.6968 |
| 17 | 0.361 | 4 | 0.6667 | 0.3333 | 0.6276 |
| 18 | 0.363 | 4 | 0.3333 | 0.6667 | 0.797 |
| 19 | 0.363 | 4 | 0.6667 | 0.3333 | 0.8747 |
| 20 | 0.363 | 4 | 0.0041 | 0.9959 | 0.9275 |
| 21 | 0.363 | 4 | 0.3333 | 0.6667 | 0.3683 |
| 22 | 0.361 | 4 | 0.9959 | 0.0041 | 0.0799 |
| 23 | 0.361 | 4 | 0.0024 | 0.9976 | 0.5744 |
| 24 | 0.390 | 4 | 0.6667 | 0.3333 | 0.0304 |

The coordination number (**Table Appendix 2**) indicates the nearest neighbour atoms surrounding each potential site, providing an understanding of the local environment that hydrogen atoms would experience upon insertion. The positions of potential hydrogen insertion sites are crucial for evaluating the material's efficiency as a hydrogen permeation barrier. The coordinates (x, y, z) in **Table Appendix 2** correspond to the normalized positions within the unit cell. We elucidated the hydrogen permeation behaviour in these structures by calculating first the electronic energy at the ground state ($E_{Structure}$) without hydrogen and $E_{Structure+H}$ with hydrogen and then the temperature-dependent insertion energies of hydrogen. $E_{Structure}$ was calculated using DFT, which estimates the total energy of electrons based on electron density. This calculation provides insights into the lowest energy state of the system, reflecting the most stable configuration of electrons under normal conditions. Understanding the electronic energy of a structure at the ground state, $E_{Structure}$, gives us essential information about the fundamental quantum interactions and stability of the material, directly influencing its reactivity and properties.

The insertion energy, $E_{Insertion}$, is crucial for understanding the energetics of hydrogen uptake into the structure matrix and is defined as the energy difference between the system with the hydrogen atom inserted into the structure lattice and the pristine structure system, mathematically represented as:

$$E_{Insertion} = E_{Structure+H} - (E_{Structure} + E_H)$$

where $E_{Structure+H}$ is the total energy (electronic energy) of the structure system with the inserted hydrogen atom, $E_{Structure}$ is the total energy of the pristine structure system without a hydrogen atom, and $E_H$ is the energy of an isolated hydrogen atom referenced to gaseous molecular hydrogen. Referencing the electronic energy of a single H atom to gaseous molecular hydrogen is a common practice for chemical binding energies, providing consistent and comparable results. While the energy of the H atom in the 1s ground state is well-defined relative to the fully ionized state (proton), this reference is less practical for calculating insertion energies in solid-state structures. Referencing gaseous molecular hydrogen allows for the inclusion of the H2 bond energy, which is more realistic for chemical processes and binding energies occurring within a material system. We then expanded our investigation to encompass the temperature dependence of the insertion energy to understand hydrogen interaction with the barrier structure lattice and assess degradation phenomena.

In electronic energy calculations, the energy minimum is sought to determine the most stable state of the system, which involves optimizing the electron configuration to minimize the total energy. Lattice vibrations (phonons) are considered separately after determining the electronic structure by calculating their contributions to the free energy and accounting for atomic movements. The DFT method focuses on calculating electronic energy, accurately describing the system's ground state. While electronic energy is typically the dominant contributor to the system's total energy, phonon contributions are essential for a complete thermodynamic picture, especially at higher temperatures. Therefore, the electronic energy calculated using DFT identifies the most stable state, and the lattice vibrations are accounted for through phonon calculations.

The formation energy of hydrogen in the structure as a function of temperature, $E_{Insertion}(T)$, was calculated to account for the thermally activated processes that could affect hydrogen solubility and mobility within the lattice. This approach acknowledges that the insertion energy landscape can change with temperature due to thermal vibrations, phase transitions or decomposition, and defect formations. The variation of the insertion energy with temperature was computed by incorporating the vibrational free energy of the hydrogen within the structural matrix and the corresponding changes in the vibrational states of the lattice, allowing all atoms to move freely without freezing any atoms, and considering lattice expansion and potential anharmonicities at elevated temperatures. By evaluating $E_{Insertion}(T)$, we aimed to characterize the temperature-activated mechanisms and the thermodynamics that govern hydrogen permeation in TiAlN, essential for predicting the material's performance in hydrogen-rich environments. The outcomes of these calculations provide insight into the thermal stability of hydrogen in the barrier structure matrix and identify temperature regimes that may either promote or inhibit hydrogen uptake.

$E_{Insertion}(T)$ for each crystalline form was deduced from their respective temperature-modified Helmholtz free energies, F(T), with the isolated hydrogen atom's free energy subtracted, as follows:

$$E_{Insertion}(T) = F_{Structure+H}(T) - F_{Structure}(T) - F_H(T)$$

where $F_{Structure+H}(T)$ is the Helmholtz free energy of the structure system with an inserted hydrogen atom at temperature *T*, $F_{Structure}(T)$ is the Helmholtz free energy of the pristine structure system at temperature *T*, and $F_H(T)$ is the Helmholtz free energy of an isolated hydrogen atom at temperature *T*. These Helmholtz free energies were derived from the phonon density of states, computed under the harmonic approximation using the phonon module within MedeA. The module performs first structural optimization by considering the atomic configurations within the created supercells, larger than the corresponding unit cells but considering lattice symmetry. Then, the phonon density of states across the entire Brillouin zone is calculated, sampling an interaction radius of approximately 10.0 Å, with asymmetric atom positions perturbed by ±0.02 Å. Finally, the Helmholtz free energy for each system at various temperatures is calculated using the phonon density of states and the partition function:

$$F(T) = E_{Structure} + F_{Vibrational}(T)$$

Where $E_{Structure}$ is the electronic (VASP) energy from DFT calculations, and $F_{vibrational}(T)$ is the vibrational contribution to the free energy at temperature T.

## Computational parameters

*Insertion energy calculations* – VASP 6 was used for spin-polarised DFT calculations. The Generalized Gradient Approximation (GGA) within the Perdew-Burke-Ernzerhof (PBE) formalism was employed for the exchange-correlation functional. The structures underwent geometry optimization calculations, which included optimizing atom positions, unit cell shape, and unit cell volume within the crystal structure. The computational setup consisted of a plane-wave cut-off energy set to 500 eV with a convergence criterion of 0.02 eV/Å. The electronic iterations were set to converge at a threshold of $10^{-6}$ eV, utilizing the Normal (blocked Davidson) algorithm alongside reciprocal space projection operators. The Brillouin zone sampling was specified at a k-spacing of 0.25 Å$^{-1}$, resulting in an 11 x 11 x 11 mesh that translates to actual k-spacings of 0.233 x 0.233 x 0.233 Å$^{-1}$ for the cubic structure and in a 7 x 7 x 3 k-point mesh corresponding to actual k-spacings of 0.214 x 0.214 x 0.250 Å$^{-1}$ for the hexagonal structure. The mesh was centred on the gamma point and maintained an odd number of points in each direction to ensure symmetry. First-order Methfessel-Paxton smearing with a width of 0.1 eV was employed to facilitate electronic convergence. We utilized version 4.0 GGA-PBE / PAW potentials for titanium (Ti_sv PAW_PBE 26Sep2005), aluminium (PAW_PBE Al 04Jan2001), nitrogen (N PAW_PBE 08Apr2002), and hydrogen (H PAW_PBE 15Jun2001) to accurately model the electronic interactions within the material. The first step of the calculations was structure minimization, which considered atom positions, unit cell shape, and unit cell volume to allow for free relaxation. Then, upon successful minimization, the symmetry of the structure was lowered to P1, and a hydrogen atom was added using the find space module in MedeA, followed by a further complete structural minimization calculation, as provided above. The position of the added hydrogen atom was done on several lattice sites. However, we focus only on the sites with the most significant space.

*Phonon calculations* – The symmetry of each structure was raised to the highest possible symmetry to reduce computational cost. The phonon module customizes the supercell size to align with the distinct lattice parameters of each structure. For example, the supercell of the hexagonal Ti$_4$AlN$_3$ structure contained 153 atoms. Then, a series of 102 single-point energy computations were executed on these augmented supercells to incorporate the effects of atomic displacements. During the optimization process, the atomic positions in the non-displaced supercell were relaxed, with the initial wave functions derived from those of previously calculated supercells.

*Migration barrier* – The Transition State Search (TSS) module was used to examine the hydrogen migration between two structures where hydrogen was positioned at adjacent vacancy sites, determined by choosing the largest vacancy sites available. The procedure utilized the Nudged Elastic Band (NEB) method, which employs a series of intermediate images to delineate transition states across the minimum energy path (MEP). Linear interpolation of initial configurations produced three intermediate states between the starting and concluding structures, utilizing a spring constant of 5 eV/Å². The identification of transition states was achieved through the Elastic Band Methods, optimized by the RMM-DIIS algorithm with a convergence threshold of 0.05 eV/Å, a maximum of 50 steps, and initial inverse Hessian matrix elements set to 0.001 Å²/eV. Electronic structure and energy calculations were conducted using VASP 6, applying the GGA-PBE functional with spin polarisation, an energy cut-off of 400 eV, and a k-point spacing of 0.5 per Å for Brillouin zone sampling, complemented by Methfessel-Paxton smearing (width 0.2 eV). The NEB iterations aimed at minimizing the MEP, with an electronic iteration convergence criterion of $10^{-5}$ eV and further refinement of transition states through the climbing image NEB method. Finally, the found transition state was minimized with more accurate VASP calculation parameters like those used for insertion energy calculations. The migration energy was then calculated as follows:

$$E_{migration} = E_{transition\ state} - E_{initial\ state}$$

The migration barrier energy of hydrogen quantifies the energy required for a hydrogen atom to migrate from one stable position to another through the crystal lattice.

*Band Structure of cubic TiAlN*

The electronic band structure for the cubic TiAlN2 structure shown in **Figure Appendix 1** was calculated to assess whether the structure has dominating metallic or semiconductor properties and to verify the plausibility of the insertion energy calculation. The band structure is shown in **Figure Appendix 4**.

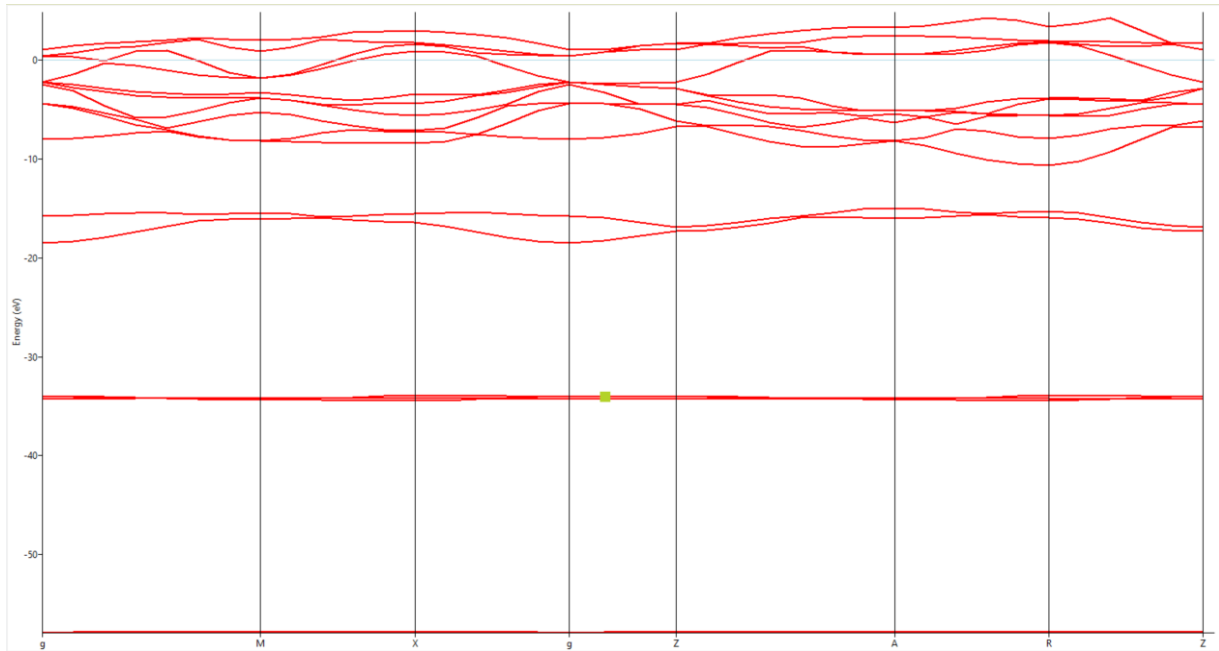

***Figure Appendix 4:*** *Electronic band structure for cubic TiAlN2 calculated using Hybrid Functional HSE06*

The structure optimization was performed using the hybrid functional HSE06 with the VASP code. A non-magnetic calculation was conducted with a plane-wave cut-off energy of 550 eV. The electronic iterations converged at $1.0 \times 10^{-6}$ eV using the damped molecular dynamics algorithm and reciprocal space projection operators. A k-point spacing of 0.25 Å$^{-1}$ was used for non-local exchange, resulting in a 7 x 7 x 7 k-point mesh. The SCF calculation used a 7 x 7 x 7 mesh, and the gamma-point-centered grid ensured an odd number of points in each direction. The linear-tetrahedron method with Blöchl corrections was applied for energy calculations. Additional settings included using initial charge density and wave functions from a previous run, and the maximum number of electronic iterations was set to 100 with up to 200 ionic steps. The electron localization function and work function (for surfaces) were also calculated.

The Work function results are as follows: upper face: -19.162 eV, lower face: -19.162 eV and average: -19.162 eV.